\documentclass[english,runningheads]{llncs}
\pdfoutput=1
\usepackage{graphicx}
\usepackage{hyperref}
\newcommand{\figref}[1]{Fig.~\ref{#1}}

\newcommand{\secref}[1]{Sec.~\ref{#1}}
\begin{document}
\title{Managing the Complexity of Processing Financial Data at Scale - an Experience Report\thanks{Preprint submitted version. Revised version to be published in the proceedings of the 10th Complex Systems Design \& Management conference (CSD\&M'19) by Springer.}}
\titlerunning{Preprint: The Complexity of Processing Financial Data at Scale}
%
\author{Sebastian Frischbier \and
Mario Paic \and
Alexander Echler \and
Christian Roth}
\authorrunning{Frischbier et al.}
%
\institute{vwd: Vereinigte Wirtschaftsdienste GmbH, Frankfurt a.M., Germany
}

\maketitle              
\begin{abstract}
	
Financial markets are extremely data-driven and regulated. Participants rely on notifications about significant events and background information that meet their requirements regarding timeliness, accuracy, and completeness. 
As one of Europe’s leading providers of financial data and regulatory solutions vwd processes a daily average of 18 billion notifications from 500+ data sources for 30 million symbols. Our large-scale geo-distributed systems handle daily peak rates of 1+ million notifications/sec.
In this paper we give practical insights about the different types of complexity we face regarding the data we process, the systems we operate, and the regulatory constraints we must comply with. We describe the volume, variety, velocity, and veracity of the data we process, the infrastructure we operate, and the architecture we apply. We illustrate the load patterns created by trading and how the markets' attention to the Brexit vote and similar events stressed our systems.

\end{abstract}
\section{Introduction}
There are many ways to decide on investments at financial markets: Intuition, psychology, tapping into social media, following recommendations of analysts and influencers, or quantitative analysis of trends and correlations. While the former approaches usually sound more intriguing as they come with a certain enigma, quantitative analysis of financial data is the prevailing tool used by market participants. 
Thus, having access to reliable, accurate, fresh, and complete information about financial markets is vital to participants.

The general public is usually more familiar with the delayed market data provided via public websites, teletext, on television or as end-of-day aggregations on the daily newspapers' financial pages.
For professional users, however, far more diverse data-driven solutions are available to support them in their decisions. These solutions give targeted insights with high information density. Examples are portfolio management systems that constantly check portfolios with individual investment strategies against real-time data to give recommendations for redeploying capital (and manage the subsequent orders directly upon approval) or market data terminals that enable experts to combine real-time market data insights with historic and reference data for in-depth analysis.

The raw data fuelling these solutions is provided as continuous streams of structured and unstructured data by various sources, e.g., exchanges, financial institutions, capital management and investment companies, and rating agencies. 
Financial data vendors and solution providers like vwd collect this data, purge, and enrich it before providing the resulting condensed information at different levels of quality of information (QoI) to subscribers. 

vwd is one of Europe's leading providers of data-driven financial solutions. We are directly connected to the majority of sources in order to process the data ourselves. Founded in 1949 as a news agency, our products and their supply
chain are nowadays completely digital. We provide solutions ranging from market data-heavy products to advisory and regulatory solutions offered as cloudbased Software-as-a-Service (SaaS) that help our customers to focus on their core
business while being compliant with regulations.
Our customers are private and public financial institutions, investment and portfolio manager, the news media in print and television as well as the general public. As a group vwd serves its customers from 14 locations in six countries. While some of our subsidiaries provide solutions directly to end users, most of our customers are intermediaries on the financial markets. Directly and indirectly 30 million users rely on our information on a daily basis to form an opinion on financial markets.

In this paper we give practical insights about the complexity of processing financial data at scale when catering to an industry that is highly regulated and competitive.
We illustrate ten challenges resulting from the volume, variety, velocity, and veracity of the data we process, the historically grown heterogeneous IT application landscape we operate, and the major regulatory constraints we have to comply with as a financial solution provider. In particular we show how regular patterns and the attention of markets to singular pivotal events reflect in demand and supply for financial data streams using our observations of the Brexit vote and the final ballot Trump vs. Clinton (2016) as examples. 

In \secref{sec:problemstatement} we identify the  challenges C1--C10 by describing the diversity of financial data ~(\secref{sec:finacialdatastreams}) and the streams processed by vwd~(\secref{sec:streamsatvwd}), the major challenges on an IT compliance level~(\secref{subsec:regulatoryconstraints}), and the heterogeneity of the IT systems we operate~(\secref{subsec:historicallygrown}). 
In \secref{sec:solutionapproach} we outline how vwd addresses the resulting complexity: we describe the infrastructure we operate and the architectural patterns we apply~(\secref{sec:architectureinfrastructure}), as well as the organizational measures we took for software development, innovation management, and compliance~(\secref{sec:solutionorganisation}). We summarize our contributions  in~\secref{sec:conclusion}.

\section{The Complexity of Processing Financial Data at Scale}
\label{sec:problemstatement}

As an international solution provider to the financial industry we do have to face several challenges -- primarily stemming from the financial data feeds we process, the regulatory constraints we have to comply with, and the historically grown heterogeneous IT systems we operate. Thus, we first give a brief introduction about those aspects of financial data that are relevant for the scope of this paper. 

\subsection{Background: Financial Data Feeds}
\label{sec:finacialdatastreams}

The umbrella term \emph{financial data} denotes a wide spectrum of unstructured and structured data with quite differing information density about \emph{financial instruments} and their issuers. Examples for financial instruments (short: \emph{instruments}) are securities/stocks, funds, futures, currencies, or indices. Please note that most but not all instruments are traded via exchanges or other platforms -- take \emph{over the counter} (OTC) securities as an example. 
Unstructured and semi-structured financial data ranges from \emph{general news} and corporate information to notifications about specific performance-related decisions by publicly traded companies (e.g., mergers, acquisitions) that must be instantly published as \emph{ad-hoc messages}. 
Structured financial data is provided as \emph{market data} at various levels of granularity and quantifies the value and prospect of a certain  instrument. In its purest and finest-granular form market (\emph{tick}) data carries information about the current trading value of a certain instrument instance (also called \emph{symbol}) at a given point in time at a specific exchange or trading platform. The most common properties used to denote this value are \emph{bid, ask, bid size, ask size, timestamp}. Market data can also contain aggregations (e.g., weighted averages) or Key Performance Indicators (KPIs) with high information density that quantify the risk/performance of an instrument in a given context (e.g., spread, betas) based on complex reasoning using historic and \emph{reference data} (metadata).

The most important drivers of complexity when processing financial data are the value of its information, how it is provisioned, and how it is represented. 

\paragraph{Quality and value of information.}  
The QoI of market data can be quantified using objective metrics such as granularity, correctness, completeness\footnote{Completeness here refers to the number of properties available per notification but also to the completeness of notifications in a stream. }, timeliness, order, and availability. For data providers QoI properties are cost drivers with costs proportionally linked to the level of required QoI. On the consumer side, the value of information (VoI) for market data with certain QoI properties depends on the purpose this information is intended to be used for by each consumer~\cite{frischbier2014managing}. Thus it is highly subjective and creates complexity when to be adhered for a large number of consumers by a data provider.  

\paragraph{Provisioning.} 
Market data is provided in a subscription-based manner as streams or bulk loads. Typical data sources are the various exchanges but also financial institutions like national banks. Data is provided as \emph{feeds} where a feed is a continuous stream of event types by a certain data source and/or market segment. A feed can be provided by a single exchange/issuer or by an intermediary financial data vendor that bundles feeds from several providers. 
Feeds are denoted as \emph{full} if data is provided without any artificial degradation in timelines or granularity. Contrastingly, a feed is \emph{delayed} if the delivery of notifications is delayed by a factor while only certain notifications are forwarded based on prioritisation in \emph{throttled} feeds. 
Consequently, \emph{aggregated} feeds deliver data at lower granularity with intra-day or end-of-day aggregations being the most prominent examples.
Subscriptions are made based on data source (e.g., feed, exchange) and quality dimensions such as granularity (e.g., tick, average), timeliness (e.g., real-time, delayed, intra-day, end-of-day), and completeness (e.g., full, throttled). Larger feeds are often split into \emph{channels} that deliver different market segments; order is always assumed.
The predominant way of delivering financial data feeds is still via direct dedicated lines using multicast and we notice that data sources even increase the use of multicast nowadays. Some feeds are also available via public internet. In most cases they have to be split up along exchanges, market segments, or instrument groups into different single feeds to compensate for the lower bandwidth and higher latency of public internet connections. 

\paragraph{Representation.} Exchanges, markets, trading platforms, and trade-reporting entities are identified using a global Market Identifier Code (MIC) standardised in ISO10383. Financial instruments are associated with an alpha-numeric identifier. For stocks/securities this can be an exchange -dependent abbreviation called the \emph{(ticker) symbol} or an international identifier such as the International Securities Identification Number (ISIN) that is standardised in ISO6166. However, different data sources may use variations of these identifiers or identifiers change over time due to mergers so that the same instrument instance is represented by different symbols based on their context. For example, the stock of Apple Inc. is known as AAPL on NASDAQ\footnote{\url{https://www.nasdaq.com/symbol/aapl}} but as APC on B\"orse Frankfurt\footnote{\url{http://en.boerse-frankfurt.de/stock/Apple-share}} while being associated with the ISIN US0378331005 as a unique identifier; moreover market data about this stock is available as AAPL.OQ (Reuters)\footnote{\url{https://www.reuters.com/finance/stocks/overview/AAPL.OQ}} or AAPL:US (Bloomberg)\footnote{\url{https://www.bloomberg.com/quote/AAPL:US}}. Hence, data must be mapped and normalized at runtime.

\subsection{Challenges Processing Financial Data at vwd}
\label{sec:streamsatvwd}

For the sake of simplicity we organize the description of concrete challenges (C1 - C4) stemming from the financial data we process at vwd along the four dimensions of Big Data~\cite{6816764}: volume, variety, velocity, and veracity.

\paragraph{(C1) Volume: The overall volume increases but varies throughout the day.}
Over the years the total volume of raw feed data processed in our ticker plant has significantly increased. In particular between 2003 and 2008 we observed an exponential increase that has turned from progressive to degressive since: the average number of daily notifications (excluding unstructured data such as news and ad-hoc messages) rose from 167 million (2003) to 937 million (2006) to 8,303 million (8.3 billion) in 2008. One explanation is the decline of floor trading at the exchanges since 2004 in favour of electronic trading. Nowadays we process on average around 18 billion notifications per day and 700,000 per second (peak rate of more than 1 million events/sec). All notifications and reference data are stored as historical prices so we can provide data for the past decades.

\begin{figure}[htbp]
	\centering
	\includegraphics[width=0.65\textwidth]{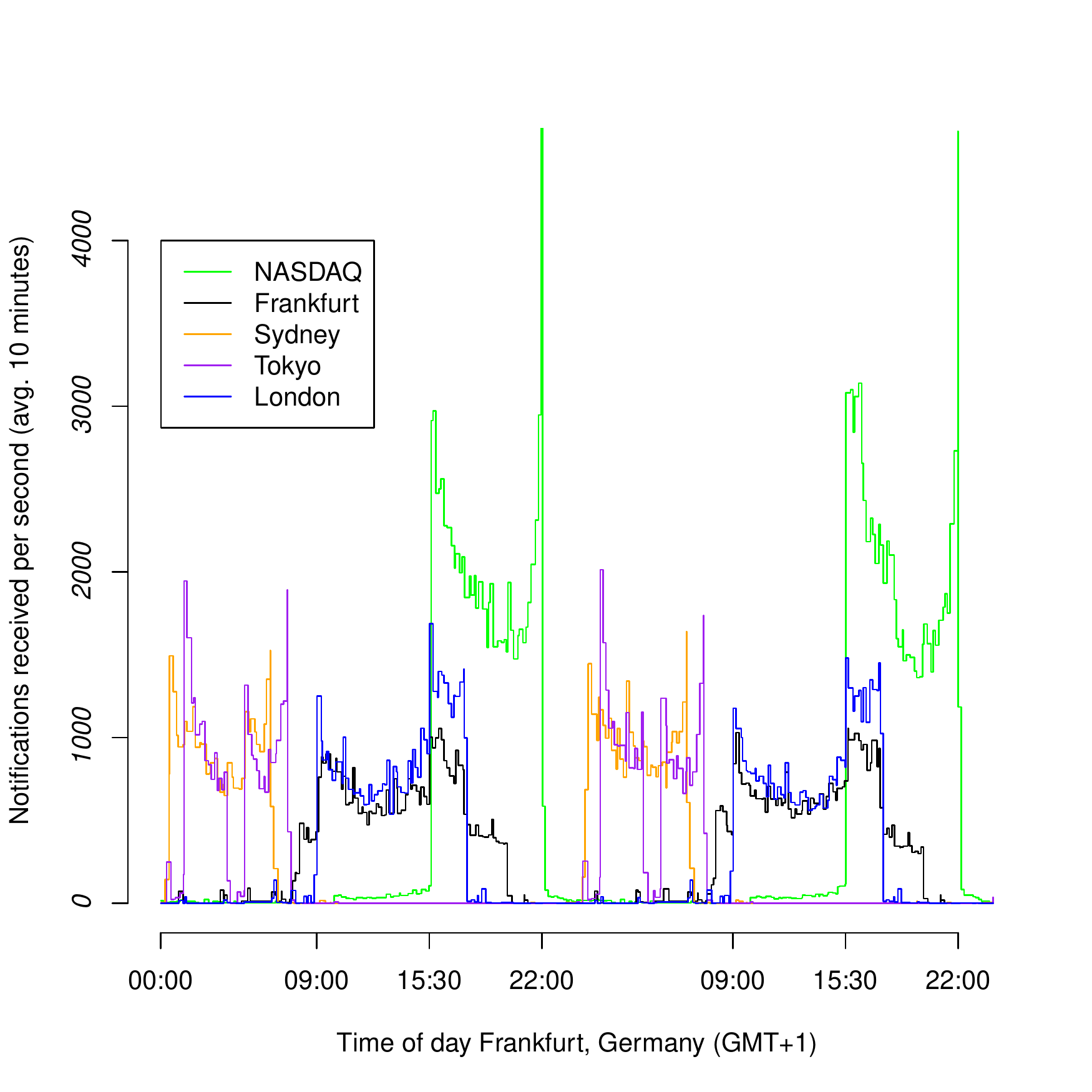}		
	\caption{Selected load received from different exchanges. Processing capacity must take the combined peak rates into account that vary massively over 24h.}
	\label{fig:exchangesovertheday}
\end{figure}

\emph{Traffic varies massively on a global and a local level throughout the day.} 
On a global level, volume differs by time zones while it also varies throughout the local trading day and per exchange. In~\figref{fig:exchangesovertheday} we show the rate of notifications received over two days for selected feeds from the exchanges London (UK, blue), Syndey (AU, orange), Tokyo (JPN, purple), NASDAQ (US, green), and B{\"o}rse Frankfurt (GER, black). Data points are averaged over 10 minute windows to smooth the plot and highlight the recurring pattern: data generated by trading on a certain exchange resembles as satchel with most activity before and after an exchange's opening time and around; there is less activity around local lunch time. Please notice the explicit 1 hour lunch break in Tokyo (purple).  We chose these exchanges as they are fairly representative for their geographic area and of comparable order of magnitude in the feed. Please note that these numbers represent only limited market segments of the actual exchange and the measured feeds might provide different products and instrument types. 

Apart from the expected peaks, announced and unannounced singular events are a separate challenge. Around the Brexit vote (June $23^{rd}$ 2016) the total volume of financial data consumed and published by us increased by 50\%. While our systems did scale to cope with this overhead, some systems of our subscribers did not. For the final ballot of the US presidential elections 2016 Trump vs. Clinton (November $8^{th}$ 2016) we dealt with an even higher volume for days.

\paragraph{(C2) Variety: Sources, formats, protocols, types, properties, notification sizes.}
We receive notifications bundled in various feeds about 120 stock exchanges, 35 futures and commodities exchanges, 180 OTC contributors and more than 500 capital management and investment companies.
The data is received and provided in custom binary or XML formats, via REST (JSON), or via bulk uploads in various file formats. 
The core processing systems at vwd process market data regarding 30 million symbols and 3 million ISINs together with reference data and unstructured data with news and ad-hoc messages. 

During a typical day 98\% of all received notifications are ticks, followed by reference data (0.16\%) and news (0.001\%). The size of these notifications (w/o news) varies between 20 bytes and 250 bytes with the average around 100 bytes. The average notification size also varies per time of day: they are larger in the morning as the \emph{open} notifications arrive, often resetting the daily statistics fields like day high/low, or total volume/turnover. 

In the evening there are fewer quotes and larger types of notifications are more common, e.g., statistics or static data.

\paragraph{(C3) Velocity: Timeliness matters for data feeds and bulk-data.} Velocity for financial data is primarily about low-latency processing of feeds and timeboxed batch-processing of bulk-data. The internal benchmark we define for low-latency processing of incoming feeds is an end-to-end latency within our ticker plant of 40~ms per notification. In addition to processing feeds and deriving complex indicators we do enrich bulk-data provided by customers such as internal ratings, portfolios, or scenario data that is fed back into their systems after we have run complex simulations. As these computations have to be finished within a given time frame they result in batch-processes with very elastic resource requirements.

\paragraph{(C4) Veracity: QoI and uncertainty of the rendered information.} For financial data veracity thus addresses QoI and its value (VoI) to subscribers that directly translates to costs. 
For our customers VoI is a function of objective data quality properties of the data they have subscribed to and their own subjective preferences; these are based on the purpose subscribers intend to use their subscribed data for. 
For example, customers using terminal solutions usually value timeliness over completeness, i.e., they prefer single notifications to be dropped completely if the information displayed on the terminal then reflects most recent events. Conversely, customers who feed the same type of notifications into their analytical systems for in-depth analysis value completeness over timeliness, i.e., they prefer a delayed complete event stream over a timely but incomplete one.

\paragraph{Summarizing challenges C1--C4.}
The characteristics of financial data pose a challenge to our processing systems: we must enrich \& push a large variety of data (C2) efficiently to our customers with as little delay and interruption as possible (C3) as its value is based on individual expectations about timeliness and completeness (C4). The large daily volume of data with its massive variations (C1) is an amplifying factor as our systems have to elastically scale to deal with peak rates that amount to several orders of magnitude of the average load.

\subsection{Challenges Regarding Compliance}
\label{subsec:regulatoryconstraints}

Financial markets are a highly regulated domain. Subsequently we do have to deal with a wide range of constraints based on European and national legislation, domain-specific regulations, and commercial license agreements for the data we process. For illustration we use examples that have a significant impact on our software development life-cycle, our application landscape, and its operations.

\paragraph{(C5) Legislation:} Most prominently, the General Data Protection Regulation (GDPR) has emphasized the importance of data locality, minimal usage, and protection for personal data while extending the scope of latter's definition. This impacts all companies active in Europe in regard to their business processes, procedures for handling data, and the software systems used. In addition to such general legislation we also do have to be aware of strict national laws applying only to the financial sector. Examples are the Swiss \emph{Bankengesetz (BankG)} or the German \emph{Gesetz {\"u}ber das Kreditwesen (KWG)} for countries we are present. They all emphasize that outsourcing organizations remain fully responsible and accountable for the services they outsource to service providers like us.

\paragraph{(C6) Regulations:}
Some services we provide rate as outsourcing agreements with organizations that are subject to further requirements by regulatory bodies. On EU level, the European Banking Authority (EBA) defines guidelines for financial institutions and financial products. Noteworthy for us in this context are in particular the refined guidelines for outsourcing arrangements~\cite{ebaoutsourcing2019}. On a national level, the German Federal Financial Supervisory Authority (BaFin), Luxemburg's Commission de Surveillance du Secteur Financier (CSSF) or the Swiss Financial Market Supervisory Authority (FINMA) define additional requirement catalogues, such as BaFin's \emph{Minimum Requirements for Risk Management for Banks}~\cite{bafinmarisk} or CSSF's guidelines regarding \emph{IT outsourcing relying on a cloud computing infrastructure}~\cite{cssfcloudoutsourcing}. All these regulations focus on the transparency, accountability, and reliability of the outsourced services and their operations. This results in additional requirements regarding reporting, change \& incident management, business continuity management (BCM), and IT security:  specific incident and compliance reporting requires further documentation and processes; identity and access management (IAM) based on the principles of least privilege and separation of duties (SoD) between operations and development has to be enforced on the physical as well as digital level and regularly reported about.

\paragraph{(C7) Licenses:} Data is the raw material of our industry. We need to maintain detailed individual license agreements with our various data sources and also with our customers about the content we process. These contracts govern service levels and penalties under which data of a certain quality is provided to us, can be redistributed by us, and consumed by our customers. This requires us to implement detailed runtime monitoring, metering, and fine-granular reporting about data provisioning and consumption down to the end user as this is essential for cost management along our digital supply chain.

\paragraph{Summarizing challenges C5--C7.}
From a compliance perspective strict requirements about transparency and availability due to regulations (C6) and license management (C7) create a tension with general legislation (C5) that enforces minimal data usage and confidentiality. This adds complexity to our software development life-cycle (SDL) and our infrastructure \& operations organization as we need to adhere to non-functional requirements such as data locality.

\subsection{Challenges Regarding IT Governance}
\label{subsec:historicallygrown}

In regard to IT governance the main challenges result from the heterogeneity of our system landscape, the need to combine pull- and push-based architectural patterns, and the scalability needed to deal with very elastic workloads.

\paragraph{(C8) Heterogeneity:} vwd has always grown both organically (new digital products and by attracting new customers) and inorganically (mergers and acquisitions, M\&A). 
On the technological level, this evolution increased the heterogeneity of IT systems to govern and operate massively over time -- in particular as our complete supply chain and product portfolio are digital. 
In general increases in heterogeneity and complexity resulting from mergers \& acquisitions are more apparent but also when growing organically new digital products tend to introduce new technology stacks while existing products rely on legacy technologies that need to be maintained. Without active governance, increasing diversity leads to technical debt along the complete technology stack and inertia. For example, at a certain point in time our Operations team had to maintain 14 different Linux distributions and versions due to acquired legacy applications that depended on specific configurations to support more than 80 legacy product families with customized versions implemented in various programming languages. 

\paragraph{(C9) Pull and push:} we need to deliver data streams to our customers, alert them about certain events and derive additional insights from the processed data on-the-fly. While this requires a push-based architectural approach~\cite{hinze2009event}, quite some of our solutions require a pull-based architectural approach as they rely on static data that needs to be pulled from data sources upon request or software solutions that are triggered by a customer's actions.

\paragraph{(C10) Scalability, Resilience, and Elasticity:} The workloads we process can vary massively in volume, variety, velocity, and veracity. Thus, we have to provide the necessary capacity in our infrastructure and enable our application landscape to elastically scale with any workload. Regulatory constraints do also require our systems to run in multiple locations for disaster resilience which requires distributed state management and additional data synchronisation.

\paragraph{Summarizing challenges C8--C10.}
In addition to general IT governance challenges stemming from historically grow system heterogeneous landscapes (C8), the data we process and the regulatory constraints we have to comply with require us to fuse pull- and push-based approaches in our architecture (C9) while operating a distributed IT system landscape that can massively scale (C10).

\section{How vwd Processes Financial Data at Scale}
\label{sec:solutionapproach}

At vwd we address the complexity of processing financial data at scale on the technological and organizational level. In this section we outline the architectural patterns we apply to our production systems, the physical infrastructure we operate across continental Europe, and the main organizational measures we take to allow for rapid development while complying with regulations.

\subsection{Technology: Modular Platform and Hybrid Infrastructures}
\label{sec:architectureinfrastructure}

On the technological level we directly address challenges C1--C4 regarding data management, C5 \& C6 regarding compliance, and C8--C10 regarding IT governance: we operate an extensive geo-distributed infrastructure and apply a modular platform approach to our production systems that combines the two complimentary paradigms of event-based systems (EBS) and service-oriented architectures (SOA) into an event-driven architecture (EDA).

\begin{figure}[!ht]
	\centering
	\includegraphics[width=0.9\textwidth]{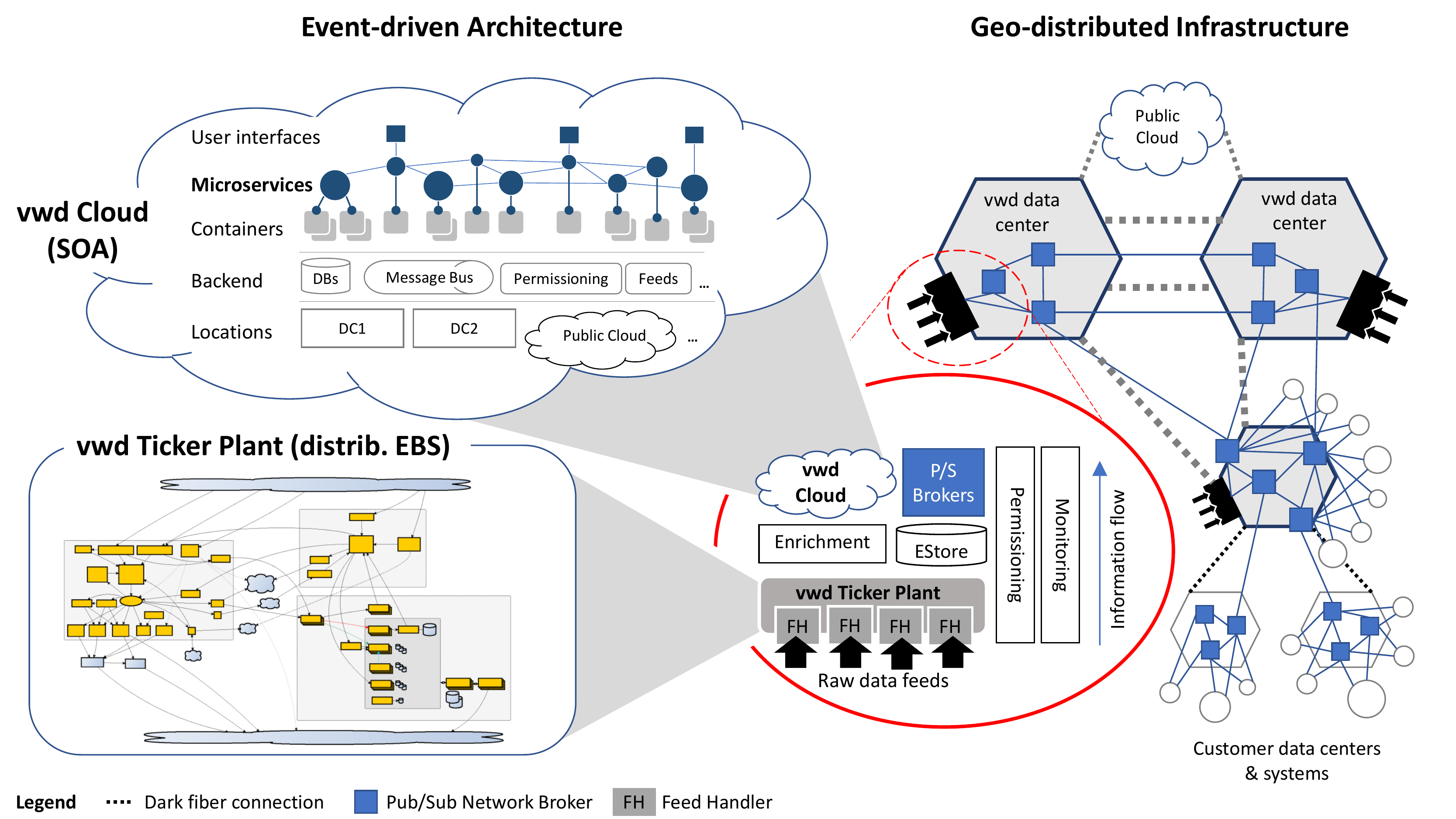}
	\caption{Big picture vwd event-driven architecture (left) with containerized service-oriented vwd Cloud (top) and vertically integrated distributed Ticker Plant (bottom); excerpt of geo-distributed physical infrastructure (right).}
	\label{fig:architecture-bigpic}
\end{figure}

\emph{Technology big picture.} The three main components of this approach are illustrated in \figref{fig:architecture-bigpic}: the target architecture of our customer-facing vwd Cloud (left, top) focuses on decoupling systems and infrastructure by using open and commercial platforms, containerization, and microservices based on Docker Swarm and Kubernetes. Close to the data sources distributed event-based systems of our Ticker Plant developed inhouse for feed processing (left, bottom) are vertically integrated with our physical infrastructure (right): as we control the complete technology stack we can customize network protocols and align software with hardware for maximum processing performance.  

\subsubsection{A Reactive Architecture.}
\label{sec:solutionarchitecture}

On the architecture level we combine push-based EBS and pull-based SOA for three reasons: 
First, to directly address challenges C1--C4 stemming from the push-based characteristics of financial data streams and those regarding licensing (C7). Second, to address the IT governance challenges C8 \& C9 by using a modular service-oriented architecture that reduce heterogeneity and allow for integrating legacy pull-based applications. Third, leverage the scalable and distributed nature of the resulting EDA to address C10 and implement disaster resilience strategies based on regulations (C6) that require a geo-distributed setup without deteriorating the quality of our services.

\emph{SOA: vwd Cloud (C8,C9,C10).} As shown on the top left side of \figref{fig:architecture-bigpic}, our customer-facing production systems are implemented by choreographing containerized microservices. This approach massively reduces the heterogeneity of our system landscape from a development point of view by encapsulating business functionality into services exposed only via a homogeneous interface and accessible via an encrypted protocol. Legacy applications are integrated by exposing their functionality to our microservice universe via a REST service as facade (e.g., Oracle's REST Data Service); the legacy system behind this facade can be optimized or replaced to reduce operational heterogeneity without impacting depending services~\cite{frischbier2012fit}. With this platform approach we can transparently combine pull- and push-based systems, seamlessly reuse and recombine existing functionality for products to exploit synergies, and reduce complexity on the infrastructure level by consolidating and standardizing systems.

\emph{EBS: vwd Ticker Plant (C1--C4,C6,C10).} As illustrated in the red explosion view (\figref{fig:architecture-bigpic}, middle) information flows from our ticker plant  (bottom) to the data-driven solutions hosted on our vwd Cloud (top); each logical component is implemented using multi-tier architectures and distributed across locations (bottom, left). Instances of feed handlers (FH) in the \emph{ticker plant} of our central processing system are subscribed to our 500+ data sources to receive, check, purge and normalize the incoming feeds. Each FH is tailored to match the specific protocol, syntax, and semantics of its feed(s) and scales horizontally with its workload. Frequent enrichment and normalization done is adding standard properties if they are not available in the original notification, e.g., total volume, open/close, or day high/low.

The normalized data is fed into our \emph{event store} (EStore) layer together with other reference data provided by customers or contributors. The event store implementation combines in-memory and relational databases. Complex events are detected in the \emph{Enrichment} component and KPIs derived by fusing real-time data with historic \& contextual data from databases.

\emph{Distributed PubSub Broker Network (C1,C3,C4,C6,C10).} We push normalized data via a \emph{distributed content-based publish/subscribe broker network} (blue overlay network in \figref{fig:architecture-bigpic}) that is deployed across locations and inter-company-borders, i.e., some brokers run on-premise in customers' locations. With this decentralised network we address disaster resilience while also minimizing traffic through application-level multicast and filtering based on fine-granular subscriptions to any of the 30 million symbols we process in our data universe. We artificially degrade QoI by throttling or delaying the streams of notifications for those subscribers that are entitled only to less-granular or less-timely data. 

\emph{Permissioning (C7).} Managing the entitlement of customers to receive data from certain feeds, metering the consumption and subsequent reporting back to data sources and authorities is handled by our central \emph{Permissioning system}.

\subsubsection{A Hybrid Multi-site Infrastructure.}
\label{sec:solutioninfrastructure}
Our systems run on a hybrid infrastructure with gigabit connectivity owned and operated by us for most parts~\cite{equinixvwdsuccess}. 

\emph{Geo-distributed physical infrastructure (C6,C10).} For regulatory and performance reasons we operate a geo-distributed physical infrastructure out of various locations as illustrated on the right hand side of  \figref{fig:architecture-bigpic}: dedicated data centres, collocation sites, managed hosting environments, and public cloud platforms. 
The overall setup is designed for disaster resilience, i.e., one unavailable site can be compensated by another. While our primary production sites are in continental Europe we are also present with our own hardware on-site at local exchanges such as Hong Kong or London for connectivity and due to local regulations.

\emph{Gigabit connectivity (C1,C3,C4,C10).} We connect our locations with redundant dedicated dark fiber gigabit lines. These also connect us with most of our contributors and customers as this is the predominant mode for exchanging financial data streams due to their volume and latency requirements; same to connect our locations to global public cloud providers. This way we can use these resources transparently within our network. Currently, one dedicated public cloud connection has 10~Gbit bandwidth but connections can be trunked.

\emph{Resource pools (C1,C3,C4,C6,C8,C10).} Within our sites we operate cascading tiers of several hundreds of physical and virtualized servers per site. While we heavily rely on virtualization for optimized resource utilization, security, resilience, and scalability we operate different resource pools based on the workload to be processed: dedicated physical servers for processing data feeds and legacy monolithic applications, virtual desktop environments for providing hosted terminals to customers, private cloud platforms for containerized systems (vwd Cloud), and public cloud environments to exploit resource elasticity for batch processing and burst-outs. Internally we use gigabit lines and software defined networks (SDN): 10~Gbit for traffic-intensive platforms and 1~Gbit for less utilized environments. 
Regarding storage we apply a mix of centralized and decentralized hyperconverged storage solutions (e.g., CEPH). Long-term storage is a crucial topic for us from a product and compliance perspective as we have to keep certain data for regulatory reasons up to ten years. Thus we combine on-site and off-site storage pools which are continuously synchronised.

\subsection{Organization: Balance Agility with Regulatory Accountability}
\label{sec:solutionorganisation}

Organizational structures and measures set the frame for technical design decisions and tooling to fully realize their potential — and vice-versa. Thus we apply a \emph{two-speed approach} for our production systems to address C6, C8, and C10 by balancing the need for reliably improving legacy systems (slow track) with fast innovation to develop new products using \emph{agile methodologies} such as SCRUM (fast track). As a key enabler for this approach we have completely restructured our organization: we reduced hierarchies, broke down technology- and location-based silos and instead created cross-location DevOps teams.       
A company-wide framework of IT policies describes how we align our processes and procedures with legal and regulatory requirements in regard to information security, service design, open source license management, and operations (C5, C6). Automation embeds these guiding principles directly in the implementation and execution of a process without the need for additional procedural descriptions. Especially Infrastructure as Code (IaC) and CI/CD pipelines guarantee the required accountability of releases and repeatability of deployments. 

\section{Conclusion}
\label{sec:conclusion}

In this paper we outlined the challenges arising from processing financial data at scale. We identified ten key challenges C1--C10 that stem from data management, IT governance, and (IT) compliance. In particular we contributed detailed insights into the processing of financial data streams along the dimensions of volume, variety, velocity, and veracity used in the context of Big Data. We outlined how these challenges are addressed at vwd on a technical and organizational level: using a geo-distributed physical infrastructure to run an event-driven platform that combines SOA and EBS paradigms to seamlessly integrate legacy systems while allowing for rapid development of new products using agile approaches.

%
%
%
\bibliographystyle{splncs04}
\bibliography{feed-processing-csdm2019-preprint}

\end{document}